\documentclass[%
 reprint,
superscriptaddress,
 amsmath,amssymb,
 aps,
prl
]{revtex4-2}

\usepackage{changes}
\usepackage{comment}
\usepackage{graphicx}
\usepackage{dcolumn}
\usepackage{bm}
\usepackage[colorlinks=true,linkcolor=blue,citecolor=blue,urlcolor=blue]{hyperref}
\usepackage[separate-uncertainty=true,multi-part-units=single,range-phrase={ - }]{siunitx}

\newcommand{\SRO}{Sr$_2$RuO$_4$}

\newcommand{\is}{\textit{in-situ}}
\newcommand{\evHs}{$\varepsilon_{\textrm{vHs}}$}

\begin{document}

\preprint{APS/123-QED}

\title{Non-Fermi liquid quasiparticles in strain-tuned Sr$_2$RuO$_4$}

\author{A. Hunter}
\affiliation{Department of Quantum Matter Physics, University of Geneva, 24 Quai Ernest-Ansermet, 1211 Geneva 4, Switzerland}

\author{C. Putzke}
\affiliation{Max Planck Institute for the Structure and Dynamics of Matter, Hamburg, Germany}

\author{F. B. Kugler}
\affiliation{Center for Computational Quantum Physics, Flatiron Institute, 162 Fifth Avenue, New York, NY 10010, USA}

\author{S. Beck}
\affiliation{Center for Computational Quantum Physics, Flatiron Institute, 162 Fifth Avenue, New York, NY 10010, USA}

\author{E. Cappelli}
\affiliation{Department of Quantum Matter Physics, University of Geneva, 24 Quai Ernest-Ansermet, 1211 Geneva 4, Switzerland}

\author{F. Margot}
\affiliation{Department of Quantum Matter Physics, University of Geneva, 24 Quai Ernest-Ansermet, 1211 Geneva 4, Switzerland}

\author{M. Straub}
\affiliation{Department of Quantum Matter Physics, University of Geneva, 24 Quai Ernest-Ansermet, 1211 Geneva 4, Switzerland}

\author{Y. Alexanian}
\affiliation{Department of Quantum Matter Physics, University of Geneva, 24 Quai Ernest-Ansermet, 1211 Geneva 4, Switzerland}

\author{J. Teyssier}
\affiliation{Department of Quantum Matter Physics, University of Geneva, 24 Quai Ernest-Ansermet, 1211 Geneva 4, Switzerland}

\author{A. de la Torre}
\affiliation{Department of Physics, Northeastern University, Boston, MA 02115, United States}
\affiliation{Quantum Materials and Sensing Institute, Northeastern University, Burlington, MA, 01803 USA}

\author{K.W. Plumb}
\affiliation{Department of Physics, Brown University, Providence, Rhode Island 02912, United States}

\author{M. D. Watson} 
\affiliation{Diamond Light Source, Harwell Campus, Didcot, OX11 0DE, United Kingdom} 

\author{T. K. Kim} 
\affiliation{Diamond Light Source, Harwell Campus, Didcot, OX11 0DE, United Kingdom} 

\author{C. Cacho} 
\affiliation{Diamond Light Source, Harwell Campus, Didcot, OX11 0DE, United Kingdom} 

\author{N.C. Plumb}
\affiliation{Swiss Light Source, Paul Scherrer Institut, CH-5232 Villigen PSI, Switzerland}

\author{M. Shi}
\affiliation{Swiss Light Source, Paul Scherrer Institut, CH-5232 Villigen PSI, Switzerland}

\author{M. Radovi\'{c}}
\affiliation{Swiss Light Source, Paul Scherrer Institut, CH-5232 Villigen PSI, Switzerland}

\author{J. Osiecki}
\affiliation{MAX IV Laboratory, Lund University, Lund SE-221 00, Sweden}

\author{C. Polley}
\affiliation{MAX IV Laboratory, Lund University, Lund SE-221 00, Sweden}

\author{D. A. Sokolov}
\affiliation{Max Planck Institute for Chemical Physics of Solids, Dresden, Germany}

\author{A. P. Mackenzie}
\affiliation{Max Planck Institute for Chemical Physics of Solids, Dresden, Germany}
\affiliation{Scottish Universities Physics Alliance, School of Physics and Astronomy, University of St. Andrews, St. Andrews KY16 9SS, United Kingdom}

\author{E. Berg}
\affiliation{Department of Condensed Matter Physics, Weizmann Institute of Science, Rehovot 76100, Israel}

\author{A. Georges}
\affiliation{Coll\`{e}ge de France, PSL University, 11 Place Marcelin Berthelot, 75005 Paris, France}
\affiliation{Center for Computational Quantum Physics, Flatiron Institute, 162 Fifth Avenue, New York, NY 10010, USA}
\affiliation{CPHT, CNRS, \'{E}cole Polytechnique, IP Paris, F-91128 Palaiseau, France}
\affiliation{Department of Quantum Matter Physics, University of Geneva, 24 Quai Ernest-Ansermet, 1211 Geneva 4, Switzerland}

\author{P.J.W. Moll}
\affiliation{Max Planck Institute for the Structure and Dynamics of Matter, Hamburg, Germany}
\affiliation{Laboratory of Quantum Materials (QMAT), Institute of Materials (IMX), École Polytechnique Fédérale de Lausanne (EPFL), Lausanne, Switzerland}

\author{A. Tamai}
\affiliation{Department of Quantum Matter Physics, University of Geneva, 24 Quai Ernest-Ansermet, 1211 Geneva 4, Switzerland}

\author{F. Baumberger}
\affiliation{Department of Quantum Matter Physics, University of Geneva, 24 Quai Ernest-Ansermet, 1211 Geneva 4, Switzerland}
\affiliation{Swiss Light Source, Paul Scherrer Institut, CH-5232 Villigen PSI, Switzerland}

\date{\today}

\begin{abstract} 
Interacting electrons can form metallic states beyond the Fermi liquid paradigm, a conceptual frontier of many-body physics mainly explored via bulk thermodynamics and transport. In contrast, the microscopics of anomalous single-particle excitations underlying non-Fermi liquid properties have largely remained in the dark. 
Here we spectroscopically map such quantum-critical excitations in \SRO{} under uniaxial pressure, an experimental challenge overcome by technical advances combining focused ion beam micro-milling with laser angle resolved photoemission.  
We show that quasiparticle excitations acquire a non-Fermi liquid scattering rate near the critical point but remain remarkably robust throughout the transition.
These experiments serve as a benchmark for the theory of anomalous metals and settle the long-standing question if quantum-critical systems host quasiparticle excitations.

\end{abstract}

\maketitle


The concept of quasiparticles -- low-energy long-lived single-particle excitations that propagate coherently -- provides the bedrock of our understanding of metals and superconductors. 
In canonical Fermi liquid metals, the decay rate of quasiparticles vanishes quadratically with their excitation energy $\omega$ from the Fermi level and with temperature $T$ resulting, via Umklapp processes, in the hallmark $\rho\propto T^2$ resistivity of Fermi liquids~\cite{Landau1937}. 
Qualitatively different metallic states have been detected in transport and thermodynamic experiments on strongly interacting systems. A prominent example is the strange metal state with $T$-linear resistivity down to the lowest temperatures reported in cuprates~\cite{Takagi1992,Proust2019,Cooper2009} and several other systems including twisted bilayer graphene~\cite{Jaoui2022,Zhang2024}. Similar phenomenology is also observed near pressure or magnetic field tuned quantum critical points in metals with otherwise well developed Fermi liquid phases~\cite{Mathur1998,Grigera2001,Custers2003,Bruin2013}. 
Non-Fermi liquid metals have long been a theoretical forefront~\cite{Varma1989,Sachdev2011,Zaanen2021,Hartnoll2022,Chowdhury2022,Hu2024}. 
However, the nature of many non-Fermi liquid states and even the existence of quasiparticle excitations in non-Fermi liquids remains controversial~\cite{Zaanen2021,Ekahana2024}. 

Angle-resolved photoemission (ARPES) is the leading experimental tool for measuring single-particle excitations. However, it is not compatible with hydrostatic pressure or high magnetic fields, the canonical tuning parameters for quantum critical phenomena. In cuprates, strange metal behavior is observed at zero field but the contribution of electron correlations to the normal state self-energy at low temperature and energy is masked by the superconducting dome and by strong signatures of electron-phonon coupling~\cite{Reber2019,Smit2024}.
The relation of transport and single-particle properties raises fundamental questions on general grounds, and is especially delicate in cuprates because of the strong momentum dependence of the 
ARPES lineshape, self energy and scattering rate.

\begin{figure*}[t]
\centering
\includegraphics[width=0.85\textwidth]{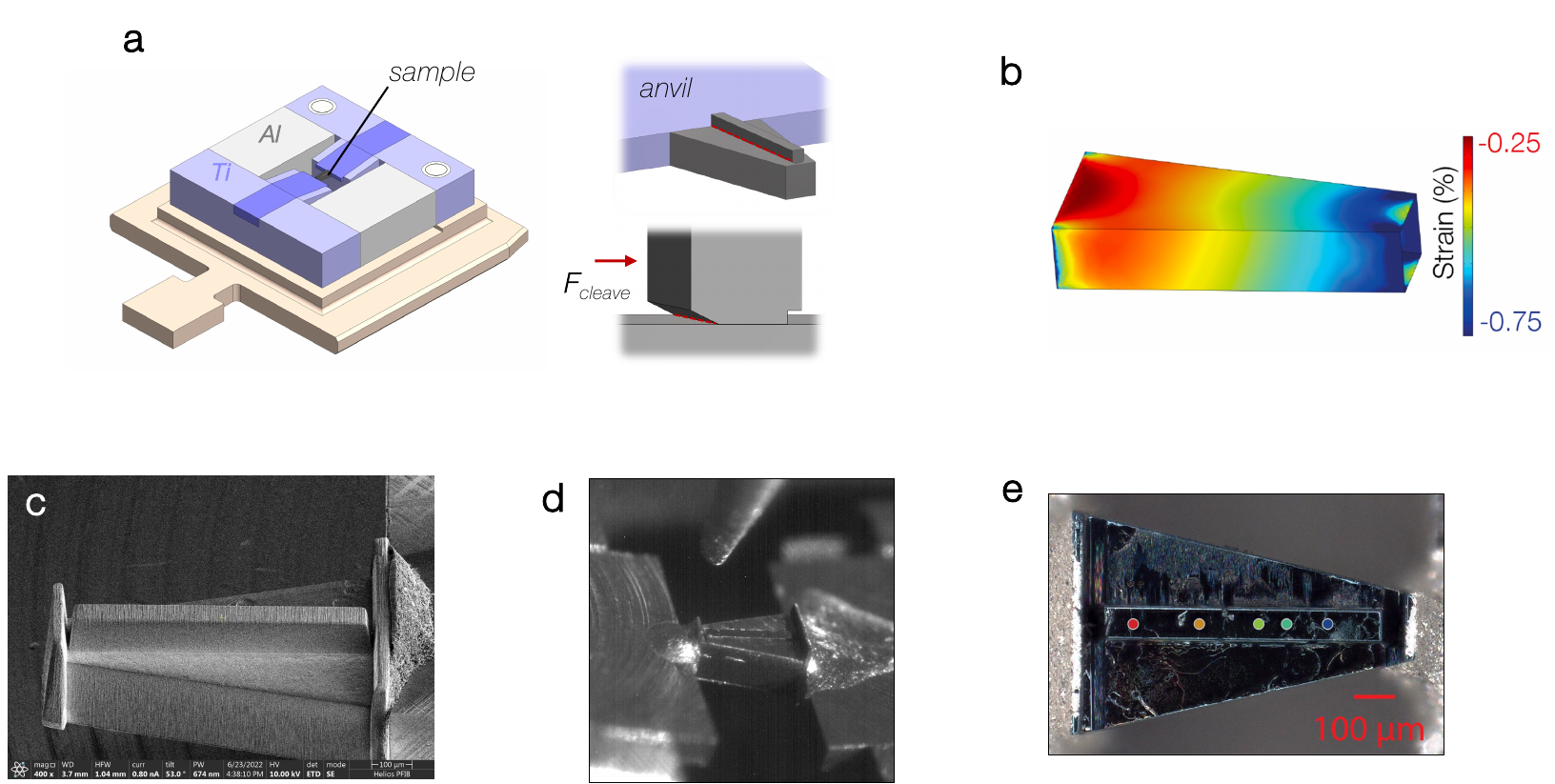}
\caption{
Controlled strain gradients in \SRO. (a) Schematic of the thermally actuated bi-metal strain cell with a close-up of the sample illustrating the undercut bar that is cleaved off in UHV before cooling the cell.
(b) Finite element analysis of the strain distribution in the tapered single crystal. The full strain cell was included in the simulation. For clarity, the figure only shows the sample.  (c) SEM image of an intermediate step in the fabrication. (d) Freshly cleaved single crystal in UHV.  (e) Optical micrograph of the FIB cut single crystal mounted in the strain cell. Different strain values are probed by moving the UV beam spot along the sample (colored dots).}
\label{fig:fab} 
\end{figure*}

The recent discovery of non-Fermi liquid transport in \SRO{} under uniaxial pressure~\cite{Barber2018} offers a new opportunity to study an anomalous metal with ARPES in a more controlled way. \SRO{} is an ultra-clean paradigmatic Fermi liquid with an unconventional superconducting phase below $T_c\approx 1.5$~K~\cite{Maeno1994,Maeno1997}. 
Strong electron correlations manifest in a dominantly local (weakly momentum dependent) self-energy renormalizing the quasiparticle dispersion of the Ru $d_{xy}$ and $d_{xz,yz}$ orbitals by factors of $\sim 5$ and $\sim 3.3$, respectively~\cite{Mackenzie1996,Tamai2019}.
Applying uniaxial pressure induces a Lifshitz transition of the largest Fermi surface sheet, commonly denoted as $\gamma$-band~\cite{Mackenzie1996}.
The Lifshitz transition changes the topology of the $\gamma$-sheet and is associated with a logarithmic van Hove singularity in the density of states crossing the chemical potential~\cite{Steppke2017,Sunko2019}.
Experiments with piezo-actuated strain cells showed that $T_c$ increases more than two-fold with strain and peaks at or near the critical strain \evHs~\cite{Hicks2014,Steppke2017}. At the same time, the temperature dependence of the resistivity changes from $T^2$ to a more singular form 
with a temperature exponent $\sim 1.5$, which is consistent with the $\rho\propto T^2\log T$ behavior expected from transport theory~\cite{Hlubina1995,Hlubina1996,Mousatov2020,Stangier2022,Herman2019,Barber2018}.

Here, we introduce a new approach to ARPES experiments under uniaxial strain to spectroscopically probe the associated non-Fermi liquid single-particle excitations. 
Previous experiments used mechanically or thermally activated strain cells in which single crystals are glued on substrates that can be strained \is~\cite{Ricco2018,Flototto2018,Sunko2019,Nicholson2021,Lin2021} or in between anvils exerting a mechanically adjustable force~\cite{Watson2017}. 
Such designs have enabled first studies of metal-insulator transitions~\cite{Ricco2018}, Lifshitz transitions~\cite{Sunko2019}, structural- and topological phase transitions~\cite{Nicholson2021,Flototto2018,Lin2021}. 
However, the reliability of this approach suffers from the difficulty to estimate the strain transmitted from substrate to irregularly shaped, cleaved single crystals.
Cleavage of typical quantum materials is essentially a stochastic process as cracks tend to propagate from defects or irregularities in the crystal shape. This prevents control over the final sample shape and thus of the strain transmitted from device to sample. Moreover, the mechanical motion of conventional strain cells often leads to irreversible strain relaxation in the epoxy or at newly formed cracks in the sample~\cite{Ricco2018}. This has resulted in generally poor predictability and reproducibility of ARPES experiments under strain.

We overcome these limitations with a conceptually simple idea illustrated in Fig.~1. 
Instead of tuning strain by mechanical motion, we precision cut samples with a defined variation in cross section $A$ resulting in a smoothly varying strain profile $\varepsilon_{xx}\propto 1/A$ upon application of a constant force. Local probing of such samples allows highly reproducible fine-tuning of the strain in any experiment with sufficient spatial resolution and control over sample position.
The strain cell used in our experiments is a simple bi-metal frame inspired by the design in Ref.~\cite{Sunko2019}. Its dimensions are chosen such that upon cooling from room temperature, the Ti anvils holding the sample induce an average strain $\Delta l/l$ of $\approx -0.6\%$ for samples with a length of $l\approx 700$~$\mu$m.
Samples are precision milled to a trapezoidal shape using a plasma focused ion beam. We define the cleavage plane by cutting a fine trench between the trapezoidal body of the sample and the central protruding bar (see Fig.~1(a) for an illustration)~\cite{Hunter2024}. This bar of $\approx 600\times50\times100$~$\mu$m is then broken off in UHV immediately prior to the experiment using a custom built precision cleaver (Fig.~1(d)).

\begin{figure*}[t]
\centering
\includegraphics[width=0.95\textwidth]{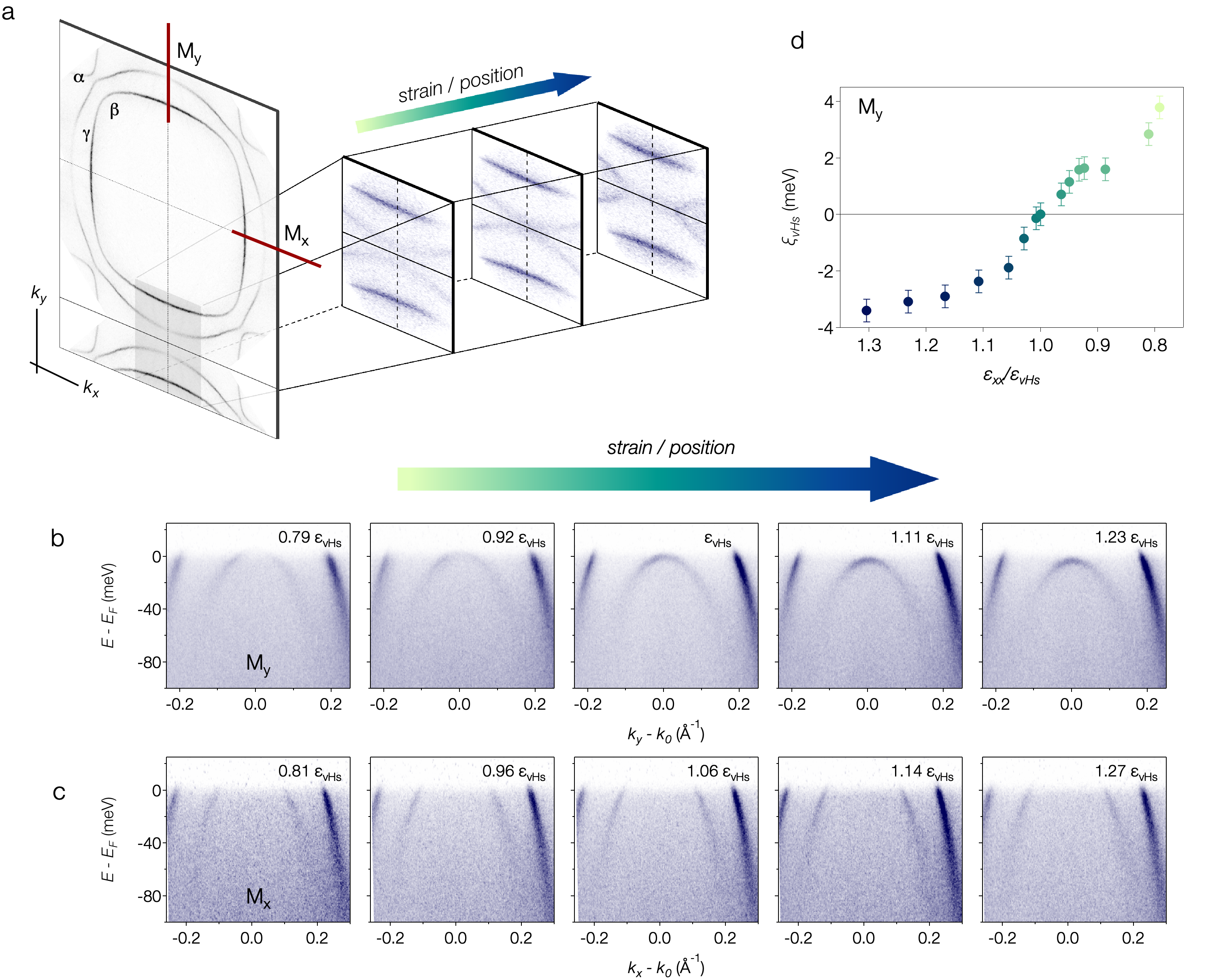}
\caption{Precision tuning of the Lifshitz transition. (a) Evolution of the laser-ARPES Fermi surface of \SRO{} with strain. The Fermi surface of unstrained \SRO{} (grey) has been measured by tilting the sample around the $(0,0) - (\pi,\pi)$ axis. Fermi surfaces of strained \SRO{} (blue) were acquired using the deflector mode of the analyser. A Lifshitz transition from a closed to an open Fermi surface of the $\gamma$ sheet is evident. 
(b,c) Energy-momentum cuts at M$_x$ and M$_y$ showing the precise tuning of the vHs across the Fermi level. (d) Energy $\xi_{\textrm{vHs}}$ of the vHs as a function of strain obtained from 2D fits of the data.
}
\label{fig:lifshitz} 
\end{figure*} 

ARPES experiments are performed with a home-built instrument combining an MBS A-1 analyzer and a narrow bandwidth 11~eV laser system from Lumeras LLC operated at 50~MHz repetition rate~\cite{Tamai2019}. 
The data are taken at $T=11$~K with $p$-polarized light and an energy resolution $\Delta E\approx 4.7$~meV. 
The laser source is focused to $\Delta x\approx 7$~$\mu$m using a spherical LiF lens mounted in UHV. This spot size corresponds to a strain resolution $\Delta\varepsilon_{xx}\approx 0.005\%$, two orders of magnitude below the critical strain \evHs.
Finite element analysis (FEA) of the full setup including strain cell, anvils, sample and the epoxies used in the assembly confirms a well-behaved, nearly linear strain profile with a gradient that varies with the aspect ratio of the sample.
Micro-focus x-ray diffraction and Raman measurements of the model system Sr$_3$Ru$_2$O$_7$ confirm the accuracy of the FEA analysis~\cite{Hunter_thesis}.
 
The strain dependence of the transport and thermodynamic properties of \SRO{} is dominated by a Lifshitz transition in the $\gamma$-band associated with the tuning across the chemical potential of a saddle point in the dispersion at M$_y = (0,\pi)$~\cite{Steppke2017,Sunko2019}.
In Fig.~2 we focus on the evolution of the band structure with strain around M$_x=(\pi,0)$ and M$_y$.
Fermi surface maps at different strain values directly confirm the transition from a closed to an open $\gamma$ sheet (Fig.~2a). The energy/momentum cuts at selected strain values in Fig.~2b,c illustrate how this transition proceeds via a gradual up/down shift of the nearly parabolic $\gamma$-band dispersion along orthogonal directions. 
The most important observation is that the single-particle excitations remain well defined at the critical strain $\varepsilon_{\textrm{vHs}}$ where the density of states at the Fermi level is singular.
This shows that quasiparticle excitations survive
in the non-Fermi liquid state detected near $\varepsilon_{\textrm{vHs}}$ in transport experiments~\cite{Barber2018}. We note that a quasiparticle picture also provides an excellent description of the unusual thermodynamic properties of \SRO{} near $\varepsilon_{\textrm{vHs}}$~\cite{Li2022,Noad2023}.

We define the van Hove energy $\xi_{\textrm{vHs}}$ as the top of the $\gamma$ band at M$_y$ (relative to the Fermi level).
In Fig.~2d, we track the evolution of $\xi_{\textrm{vHs}}$ with strain from two-dimensional fits of the measured quasiparticle dispersion.
This demonstrates sub-meV precision in placing the vHs in our experiments. Intriguingly, the strain dependence of $\xi_{\textrm{vHs}}$ appears to be more pronounced near $\varepsilon_{\textrm{vHs}}$. This is not expected from first-principles calculations of the bare band structure~\cite{Barber2019} and is opposite to the correlation-induced pinning of the vHs at the chemical potential discussed in the context of the van Hove scenario of cuprate high-T$_c$ superconductors~\cite{Markiewicz1997}.
On the other hand, the behavior observed in our experiments
is qualitatively consistent with the lattice softening observed near $\varepsilon_{\textrm{vHs}}$~\cite{Noad2023} which is not taken into account in the FEA analysis used to calibrate the strain axis.

\begin{figure*}[t]
\centering
\includegraphics[width=1\textwidth]{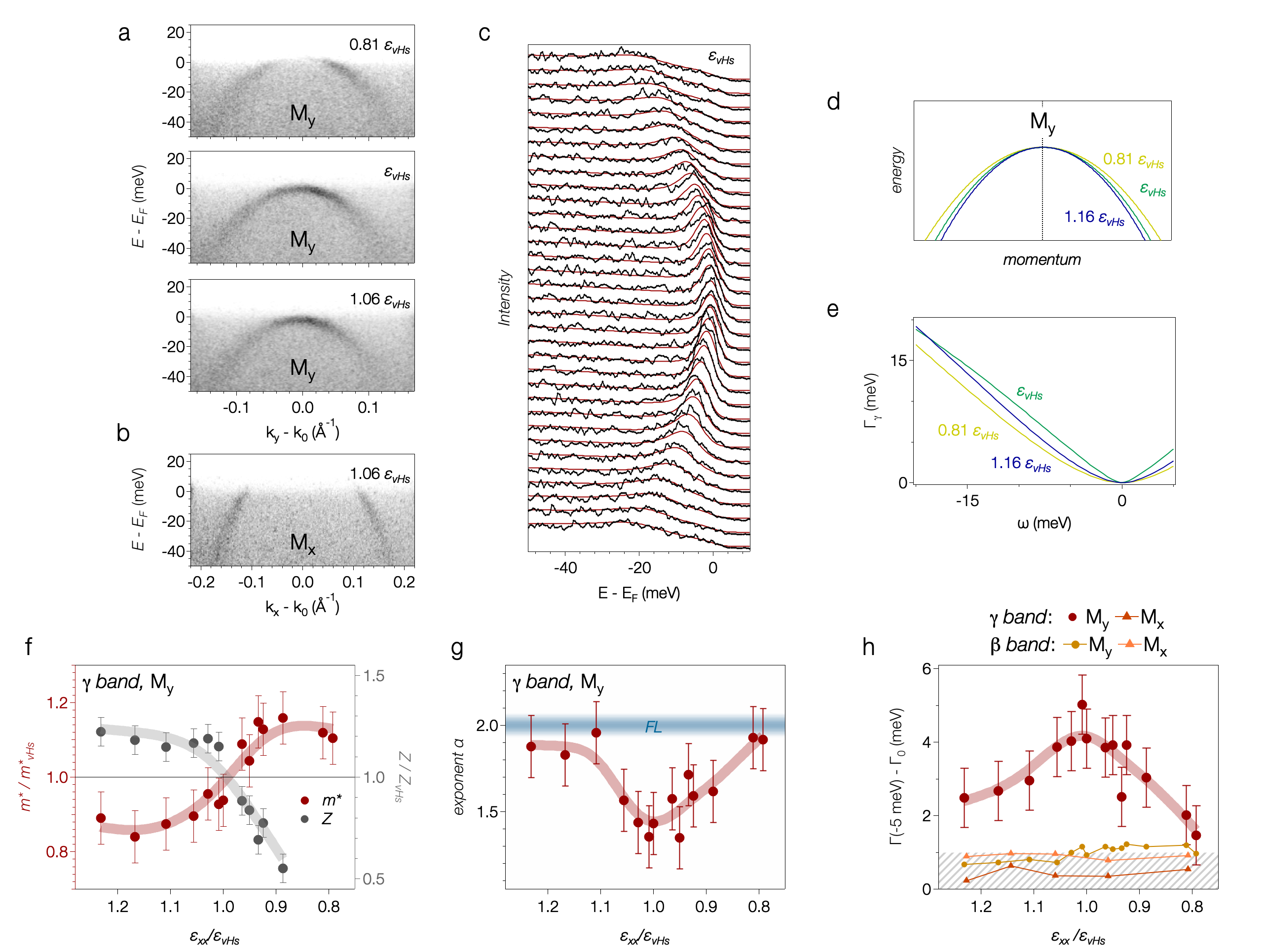}
\caption{
Non Fermi liquid quasiparticles. (a) Spectral function of the $\gamma$-band at M$_y$ near critical strain. (b) $\gamma$-band at M$_x$.
(c) Stack of EDCs (black) for \evHs{} together with a 2D fit (red).
(d) Selected $\gamma$-band dispersions $\xi_{\gamma}-\xi_{\textrm{vHs}}$ near critical strain obtained from 2D fits in the interval [-10,10]~meV.
(e) Representative $\gamma$-band line widths $\Gamma_{\gamma}(\omega)$ at M$_y$ near critical strain.
(f) Strain evolution of the mass enhancement $m^{*}/m^{*}_{\textrm{vHs}}$ and relative spectral weight $Z/Z_{\textrm{vHs}}$ of the $\gamma$-band at M$_y$. 
(g) Exponent of the $\gamma$-band quasiparticle scattering rate at the Fermi level. The blue bar indicates the Fermi liquid exponent 2.
(h) Quasiparticle scattering rates at $\omega=-5$~meV (full width half maximum). The gray shading indicates the sensitivity limit of the experiment.
}
\label{fig:img_plots} 
\end{figure*} 

The high-resolution energy-momentum cuts tracing the Lifshitz transition allow us to analyze the strain evolution of the electron self-energy $\Sigma=\Sigma'+i\Sigma''$ encoding the many-body interactions.
In Fig.~3a,b, we compare the $\gamma$-band spectral function at selected strain values. A visual inspection of the data already reveals important features. 
We first note that the $\gamma$-band dispersion remains remarkably robust. Our data at the critical strain \evHs{} does not show any signs of the extended flat region seen in early renormalization group calculations of van Hove systems~\cite{Irkhin2002} and in ARPES experiments on highly doped epitaxial graphene~\cite{Link2019}.
Instead, we find that the coherent spectral weight at M$_y$ increases monotonically with strain. This is unlikely a simple matrix element effect as neither the experimental geometry nor the orbital character of the bands are affected by strain. Instead, the increasing weight points to a monotonic increase in the quasiparticle residue $Z$ with increasing strain. 

Comparing the spectral functions at similar strain (1.06 \evHs) at the orthogonal M$_x$ and M$_y$ points, we further notice that the quasiparticle amplitude (peak height) decays much more rapidly with increasing energy near M$_y$. 
This points to marked changes in the imaginary part of the self-energy between the orthogonal M$_x$ and M$_y$ points.
The coherent spectral weight $Z$ can be approximated by the product of the quasiparticle peak amplitude and width. Assuming that $Z$ varies slowly in the small momentum windows of these cuts, 
the rapid decay of the amplitude at M$_y$ with increasing energy implies an increased slope of the imaginary part of the self-energy near the vHs. This points to deviations from the canonical form $\Sigma''\propto \omega^2$ of Fermi liquids.

We quantify these effects from 2D fits to the data that include the $\beta$ and $\gamma$ bands
(the $\alpha$-band lies outside the energy-momentum range of interest).
Two-dimensional fits allow for an accurate and model-independent treatment of resolution effects, which is essential at the low energies of interest here. 
We start by writing the photoemission intensity in the standard form \mbox{$I = \left[\left(\sum_{\nu}|M_{\nu}|^2 A_{\nu} + B\right)\cdot f\right]\ast R$}. Here, $\nu=\beta,\gamma$ is a band index, $M_{\nu}$ the photoemission matrix element, $B$ a smooth background, $f$ the Fermi function and $R$ the resolution function described by a 2D Gaussian. 
As justified in the supplementary text, 
we write the spectral function $A_{\nu}$ in terms of the renormalized quasiparticle dispersion $\xi_{\nu}(k)$ (as obtained from a fit to the data) and a locally momentum-independent quasiparticle residue $Z_{\nu}$ and quasiparticle line width $\Gamma_{\nu}(\omega)=-2Z_{\nu}\Sigma''_{\nu}(\omega)$ (full width half maximum). 
\begin{equation}
    A_{\nu}(k,\omega) = \frac{1}{\pi}Z_{\nu} \frac{\Gamma_{\nu}/2}{[\omega-\xi_{\nu}(k)]^2 + (\Gamma_{\nu}/2)^2},
    \label{AkE}
\end{equation}
where $\omega=E-E_F$. 
Momentum space differentiation is included by fitting $Z_{\nu}, \xi_{\nu}$ and $\Gamma_{\nu}$  independently around M$_x$ and M$_y$. To capture possible non-Fermi liquid behavior in the scattering rate $\Gamma=\hbar/\tau$, we parametrize the quasiparticle line width as $\Gamma(\omega)=\Gamma_0 + c|\omega|^{(\alpha-b|\omega|)}$. Here, $\Gamma_0$ is a small impurity scattering contribution which is typically around 1~meV in our data. This purely empirical form can capture Fermi liquid, marginal Fermi liquid or intermediate behaviors with a minimal set of parameters. It also gives a good description of the transition from $\Sigma''\propto \omega^2$ at low energy to the linear or sub-linear behavior at a few 10~meV observed in DMFT calculations of unstrained \SRO{} (see supplementary materials).

Fig.~3c shows that this approach provides a good fit of the ARPES intensity over the full energy--momentum range of interest. This validates the assumption of a negligible momentum dependence of the self-energy in the vicinity of M$_y$.
We first focus on the quasiparticle effective mass $m^{*}$, which serves as a proxy for the strength of correlations. Since we are primarily interested in the mass enhancement of the $\gamma$-band near M$_y$ we define a local effective mass $m^{*}_{k_0}$ from expanding the quasiparticle dispersion as $\xi_{\gamma}(k)=\xi_{\textrm{vHs}}-(k-k_0)^2/2m^{*}_{k_0}$. 
Note that this mass differs from the thermodynamic mass (interacting density of states), which diverges in 2D even in the absence of interactions as the vHs is tuned across the Fermi level. 
Indeed at the DFT level $m^{*}_{k_0}$ is largely independent of strain as shown in supplementary materials, Fig.~S6.

Fig.~3f displays the quasiparticle effective mass $m^{*}_{k_0}$ (hereafter, $m^{*}$) determined from fits of the ARPES data over a small energy range of $[-10,10]$~meV, chosen to minimize systematic errors from the non-parabolicity of $\xi_{\gamma}$.
This reveals a variation of $m^{*}$ of $\approx 25\%$ within the strain range probed in our experiment, far greater than expected from band structure effects. We thus attribute the strain dependence of $m^{*}$ mainly to many-body effects.
Notably, our analysis shows that the mass enhancement does not peak at the van Hove strain. Rather, $m^{*}$ decreases monotonically with increasing strain pointing to reduced correlations at high strain.

The quasiparticle residue $Z$ provides an alternative measure of the strength of correlations. Absolute values of $Z$ cannot be determined from photoemission data alone. However, the fits readily yield $|M_{\nu}|^2 Z_{\nu}$. In Fig.~3f we trace $Z_{\gamma}/Z_{\textrm{vHs}}$ assuming that $|M_{\gamma}|^2$ does not vary significantly with strain. This reveals a monotonic increase of $Z_{\gamma}$ with strain, qualitatively consistent with the behavior of $m^{*}$. Indeed, for a self-energy depending on energy only one has $Z^{-1}=m^{*}/m_{\textrm{bare}}$. The consistent trend between $m^*/m$ and $1/Z$ suggests that the momentum dependence of the self-energy within a given patch of the Brillouin zone is rather weak, as our fits also indicate - see supplementary text for a more detailed discussion.
In supplementary materials, we further show that $Z_{\beta}$ depends only weakly on strain, suggesting a unique behavior of correlations in the $\gamma$-band at M$_y$. 

We determine the quasiparticle line widths $\Gamma(\omega)$ from fits that include the $\beta$ and $\gamma$-band and run over a larger energy range of $[-50,10]$~meV. 
Here, we focus on the low-energy behavior of the $\gamma$-band at the vHs point M$_y$, which is most strongly affected by van Hove physics. Representative widths $\Gamma(\omega)$ for different strains are shown in Fig.~3e.
This immediately reveals anomalous behavior. At the critical strain, $\Gamma(\omega)$ clearly deviates from the canonical $\omega^2$ form and approaches -- though does not fully reach -- singular $|\omega|$ behavior. 
We quantify this behavior with the exponent $\alpha$ in the parametrization $\Gamma(\omega)=\Gamma_0 + c|\omega|^{(\alpha-b|\omega|)}$ used in the fits.
We find that deviations from Fermi liquid behavior are systematic with exponents $\alpha=\SIrange{1.3}{1.6}{}$ for a range of strains near \evHs{} (Fig.~3g). 
Fermi liquid behavior with $\alpha$ approaching 2 is recovered only for the highest and lowest strains studied in our experiment. 
We interpret the non-Fermi-liquid exponents as an average over a few meV from the Fermi level. Our experiments do not exclude even more singular behavior at very low energy and/or lower temperature.

Notably, the anomalous $\gamma$-band self-energy  coincides with a strong enhancement of the scattering rate at finite energy.
Fig.~3h shows that at M$_y$, $\Gamma_{\gamma}$ reaches $\sim 4$~meV at $\omega=-5$~meV and appears to peak near \evHs{} with a slow decrease towards high and low strain. This is in sharp contrast with all other scattering rates, which we find to be largely independent of  
strain and near or below our detection limit of $\sim 1$~meV at $\omega=-5$~meV.

The Lifshitz transition in critically strained \SRO{} can be seen as a topological quantum phase transition. Our results establish that quasiparticles remain well defined throughout this transition but acquire an anomalous scattering rate. 
Theoretical studies of generic quasiparticle bands undergoing a Lifshitz transition find a singular self energy $\Sigma''\propto |\omega|$ at van Hove filling at the momentum $k_{\textrm{vHs}}$ of the saddle point singularity and $\Sigma(\omega)\propto |\omega|^{3/2}$ for generic momenta away from $k_{\textrm{vHs}}$~\cite{Gopalan1992,Pattnaik1992}. 
However, the extension in energy, momentum and temperature of the $|\omega|$-linear behavior is not known.
Our experiments are broadly consistent with these predictions but imply that a possible $|\omega|$-linear regime is narrowly confined. We further note that $\Sigma''(\omega)\propto |\omega|$ corresponds to a logarithmically vanishing quasiparticle residue $Z\rightarrow 0$. The observation in our work of quasiparticle-like excitations with significant weight at the critical strain thus provides further evidence for self-energies that deviate from an $|\omega|$-linear behavior at the energies and temperatures of a few meV and a few K probed in our experiment.

Non Fermi-liquid behavior was previously found in resistivity measurements~\cite{Barber2018}.
We note, however, that transport scattering times and exponents differ in general from the single particle self-energy.
Electrical transport in a van Hove system is dominated by large-angle scattering between "hot" states at $k_{\textrm{vHs}}$ and "cold" parts of the Fermi surface at generic momenta away from the vHs~\cite{Mousatov2020,Stangier2022}. In contrast, the single-particle scattering rate probed by ARPES is likely dominated by small angle scattering of "hot" states near the van Hove point~\cite{Gopalan1992,Pattnaik1992,Mousatov2020,Stangier2022}.

Our analysis further demonstrates that the $\gamma$-band self-energy acquires a pronounced momentum differentiation between M$_x$ and M$_y$ near the critical strain.
Anisotropic scattering rates are well established in cuprates and are generally expected in presence of strong antiferromagnetic (AF) spin fluctuations or electron-phonon coupling~\cite{Lohneysen2007,Dahm2009,Deveraux2004}. 
The anisotropy in critically strained \SRO, revealed here, is likely of different origin.
Previous work established that the self-energy of unstrained \SRO{} is dominated by local correlations and has only small contributions from electron-phonon interaction or coupling to AF spin fluctuations~\cite{Tamai2019,Abramovitch2023}. 
We note that magnetic order, possibly in the form of a spin-density wave, has been observed in highly strained \SRO~\cite{Grinenko2021}. Our experiments, however, show a peak in the scattering rate at the critical strain and a decrease towards the magnetic phase at high strain, opposite to the expectation for spin-fluctuation scattering. 
Our results thus suggest that $k$-space differentiation is intrinsic to the many-body physics of systems with a vHs at the Fermi level and does not require nearby ordered states.

We finally highlight the fact that the strain evolutions of scattering rates and effective masses are different in our experiments. We find that $\Gamma$ peaks at the critical strain while $m^{*}$ decreases monotonically with increasing strain. This dichotomy is not expected from generic models and is not seen in previous theoretical work~\cite{Gopalan1992,Hlubina1995,Hlubina1996,Irkhin2002,Herman2019,Stangier2022}. 
It is likely related to specific properties of \SRO, especially the particle-hole asymmetry of the $xy$ density of states.
Indeed, the monotonic decrease of $m^{*}$ is reproduced in our high-precision single-site dynamical mean-field theory (DMFT) calculations of strained \SRO. As shown in supplementary materials, these calculations are in fair agreement with the experimental effective masses. Further, they reproduce the experimental trend in $\Gamma(\varepsilon)$ as well as deviations from $\Gamma\propto\omega^2$ near \evHs. We note, however, that quantitative agreement with experiment is not expected within the single-site DMFT approximation of a momentum-independent self-energy.

In summary, we spectroscopically trace a Fermi liquid through a topological quantum phase transition. 
We find that quasiparticles acquire an anomalous scattering rate near the critical point but show no sign of a singularity in either the mass enhancement or the residue.
The technical innovations underlying our results mark a milestone for surface sensitive spectroscopy in precision shape controlled crystals of complex materials. Within the burgeoning field of designer quantum materials, these methods will enable new insights into the quasiparticle spectrum of confined, distorted, and nano-patterned crystalline matter.
Here, they uncovered concise evidence for quasiparticles at a topological quantum critical point, demonstrating how the remarkable robustness of the quasiparticle concept extends well into the most singular environments of quantum criticality.

\begin{acknowledgements}
We thank J. Schmalian and P.D.C. King for discussions.
The experimental work was supported by the Swiss National Science Foundation (SNSF) grants 165791, 184998, 189657. 
We acknowledge Diamond Light Source for time on Beamline I05 under Proposal SI38224 and the Paul Scherrer Institut, Villigen, Switzerland for provision of synchrotron radiation beamtime at the SIS beamline of the SLS.
We acknowledge MAX IV Laboratory for time on Beamline Bloch under Proposal 20220192. Research conducted at MAX IV, a Swedish national user facility, is supported by the Swedish Research council under contract 2018-07152, the Swedish Governmental Agency for Innovation Systems under contract 2018-04969, and Formas under contract 2019-02496.
Work performed at Brown University was supported by the U.S. Department of Energy, Office of Science, Office of Basic Energy Sciences, under Award Number DE-SC0021265.
The Flatiron Institute is a division of the Simons Foundation. 

\end{acknowledgements}

\providecommand{\noopsort}[1]{}\providecommand{\singleletter}[1]{#1}%

\end{document}